\documentclass[12pt]{iopart}
\usepackage{latexsym}
\usepackage{amsfonts}
\usepackage{epsfig}
\usepackage{setspace} 
\usepackage{graphicx}
\usepackage{dcolumn}
\usepackage{bm}
\begin{document}

\title{Pressure Effects in Supercooled Water:
Comparison between a 2D Model of Water and Experiments for Surface
Water on a Protein.}

\author{
Giancarlo Franzese$^1$,
Kevin Stokely$^{2}$, 
Xiang-qiang Chu$^{3}$,
Pradeep Kumar$^{2,4}$, 
Marco G. Mazza$^{2}$, 
Sow-Hsin Chen$^{3}$,
and H.~Eugene Stanley$^2$}

\address{$^1$Departament de F\'{\i}sica Fonamental, Universitat de Barcelona,
Diagonal 647, 
08028 Barcelona, Spain\\
$^2$Center for Polymer Studies and Department of Physics, 
Boston University,
Boston, MA 02215 USA\\
$^3$Department of Nuclear Science and Engineering, Massachussetts
Institute of Technology, Cambridge, MA 02139 USA\\
$^4$Center for Studies in Physics and Biology, Rockefeller
  University, New York, NY 10021 USA
}
\ead{gfranzese@ub.edu}

\begin{abstract}

Water's behavior differs from that of normal fluids, having more than
sixty anomalies.  Simulations and theories propose that many of these
anomalies result from the coexistence of two liquid phases with
different densities.  Experiments in bulk water confirm the existence of
two local arrangements of water molecules with different densities, but,
because of inevitable freezing at low temperature $T$, cannot ascertain
whether the two arrangements separate into two phases.  To avoid the
freezing, new experiments measure the dynamics of water at low $T$ on
the surface of proteins, finding a crossover from a non-Arrhenius regime
at high $T$ to a regime that is approximately Arrhenius at low $T$.
Motivated by these experiments, Kumar {\it et al.}
[Phys. Rev. Lett. {\bf 100}, 105701 (2008)] investigated, by Monte Carlo
simulations and mean field calculations on a cell model for water in
two dimensions (2D), the
relation of the dynamic crossover with the coexistence of two liquid
phases. They show that the crossover in the orientational correlation
time $\tau$ is a consequence of the rearrangement of the hydrogen bonds
at low $T$, and predict that:

\begin{itemize}

\item[(i)  ] the dynamic crossover is isochronic, i.e. the value of the
  crossover time $\tau_{\rm L}$ is approximately independent of
  pressure $P$;

\item[(ii) ] the Arrhenius activation energy $E_{\rm A}(P)$ of the
  low-$T$ regime decreases upon increasing $P$;

\item[(iii)] the temperature $T^*(P)$ at which $\tau$ reaches a
fixed macroscopic time $\tau^*\geq \tau_{\rm L}$ decreases upon
increasing $P$; in particular, this is true also for the crossover
temperature $T_{\rm L}(P)$ at which $\tau=\tau_{\rm L}$.

\end{itemize}

Here, we compare these predictions with recent quasielastic neutron
scattering (QENS) experiments performed by X.-Q. Chu {\it et al.}  on
hydrated proteins at different values of $P$. We find that the
experiments are consistent with these three predictions.

\end{abstract}

\maketitle

\section{Introduction}

Water has many anomalies compared to normal liquids \cite{sitges}.
Experiments show that its thermodynamics fluctuations and response
functions, such as the isobaric specific heat $C_P$ or the magnitude of
the isobaric thermal expansion coefficient $\alpha_P$, largely increase
when temperature is decreased \cite{Angell73}.  These anomalies are more
pronounced in the supercooled liquid state, below 0$^\circ$C. This
state, metastable with respect to ice, can be observed at temperatures
as low as $-47^\circ$C in plants \cite{debenedettibook}, $-41^\circ$C in
laboratory at atmospheric pressure \cite{Cwilong1947} and $-92^\circ$C
at 2 kbar \cite{Kanno1975}.

The anomalies have been interpreted on the basis of models that propose
different scenarios. The scenarios can be divided in two main
categories: $(a)$ those scenarios that include the coexistence at low
$T$ of two liquids with different densities, and $(b)$ a scenario in
which water forms local regions of different densities, but does not
separate into two phases.

\subsection{Scenarios with coexistence of two liquids}

\noindent {\it The liquid-liquid critical point scenario}.  In 1992
Poole {\it et al.}~\cite{Poole}, on the basis of molecular dynamic
simulations for ST2- and TIP4P-water, proposed that supercooled water
separates into two liquid phases with different densities below 200~K
and above 150~MPa. The low-density liquid (LDL) form of water and the
high-density liquid (HDL) coexist along a first-order phase separation
line with negative slope in the pressure-temperature $P$--$T$ phase
diagram and terminate in a liquid-liquid critical point (LLCP).  The
LLCP scenario has been confirmed by simulations with various water
models~\cite{Brovchenko}.

The occurrence of the LLCP has been rationalized with different
theoretical models by assuming (a) the anticorrelation between energy
and volume in the H (hydrogen) bonds formation and (b) the possibility
of forming different kinds of H bonds.  An example is the thermodynamic
free energy model of Poole {\it et al.}\ that hypothesizes the formation
of strong H bonds when a geometrical condition on the molar volume is
satisfied \cite{psgsa}.

Another example is the microscopic cell model introduced in
Ref.~\cite{fs,fsPhysA} and studied in detail in Ref.~\cite{fms,fs2007},
in which the correlation between the H bonds, due to the O-O-O
interaction, is incorporated. In this model the liquid-liquid (LL) phase
transition is due to the tetrahedral ordering at low $T$ and $P$ of the
H bonds, as explained in the following section.

The LLCP is predicted to lie in a region of the $P$--$T$ phase diagram
where the freezing of bulk water is inevitable. Therefore, direct
experimental verification of the LLCP scenario is challenging.  However,
Soper and Ricci in 2000 showed with neutron diffraction measurements
that the local arrangement of water molecules changes up to the second
shell, increasing the local density when $P$ is increased from 26~MPa to
400~MPa at -5.15$^\circ$C \cite{Soper-Ricci00}.  This structural change
was initially observed by varying $T$ from about 263~K to 313~K in x-ray
structure factor experiments for heavy water D$_2$O in 1983
\cite{BCT83}. The data show that by decreasing $T$ the average O-O-O
angle increases toward the tetrahedral angle 109.47$^\circ$
\cite{BCT83}. This result has been reaffirmed by Ricci {\it et al.} in a
recent experiment \cite{ricci}.

\bigskip

\noindent {\it The critical point free scenario}.  Another scenario that
hypothesizes a liquid-liquid (LL) phase transition has been considered
recently \cite{angell2008}. In this scenario the LL first order phase
transition extends to negative $P$ and merges the liquid spinodal, but
without a critical point.  A rationalization of this scenario has been
recently proposed on the basis of the microscopic cell model for water
\cite{fs,kevin-condmat}, showing that this scenario has a liquid
spinodal that reenters from negative to positive $P$
\cite{kevin-condmat}.

\bigskip

\noindent {\it The stability limit scenario.}  The reentrant spinodal
discussed in Ref.~\cite{kevin-condmat} was earlier proposed in the
stability limit scenario \cite{Speedy82} as the origin of the anomalies
of water. Although in the initial formulation of the stability limit
scenario the occurrence of the LL phase transition was not hypothesized,
it was successively introduced for thermodynamic consistency
\cite{debenedetti_dantonio88} and found also in the thermodynamic free
energy model of Poole {\it et al.} \cite{psgsa}.

\subsection{Scenario without coexistence of two liquids}
\noindent {\it The singularity free scenario}. This scenario assumes
that the H-bonds linking molecules are uncorrelated. Under this
hypothesis, water anomalies are the effect of the negative
volume-entropy cross fluctuations \cite{Sastry96,StanleyTeixeira80} and
the large increase of response functions seen in the experiments
represents only an apparent singularity, due to local density
fluctuations.  The regions with different local density do not form
separate phases.  A pressure increase gives rise to a sharp, but
continuous, increase of density, as in the supercritical region of the
LLCP scenario. The continuous structural change is found also in {\it ab
  initio} water simulations \cite{SGGabinition00} at very high pressure,
$P=10^4$~MPa, and $T=600$~K.  This scenario is recovered by the
microscopic cell model \cite{fs} in the limiting case of no
cooperativity among the H bonds.

The singularity-free scenario, in the region accessible by experiments,
reproduces the same phase diagram as the scenarios with coexistence of
two liquids. Therefore, it is interesting to understand if there are
oberservable differences among these scenarios. In particular, recent
experiments and simulations, described in the next sections, have
analyzed the case of the dynamics of water surrounding proteins or
confined in nanopores, interpreting the experimental results within the
context of the different scenarios.

\section{Hydrated proteins}

Recent experiments on surface water and water hydrating proteins
\cite{Chen,Mamontov,Jansson} have shown that liquid water exists at
temperatures as low as $-113^\circ$C \cite{Mallamace} at ambient
pressure.  At these extremely low temperatures interesting dynamical
phenomena occur \cite{Swenson,slaving,PRLcomments1,PRLcomments2},
suggesting a possible relation for the dynamics of the biological
macromolecules with that of the surrounding water
\cite{Jansson,Brovchenko_bio}.

At low $T$, proteins exist in a (``glassy'') state with no
conformational flexibility and with very low biological activity. For
hydrated proteins above about 220~K, the flexibility is restored,
allowing the sampling of more conformational sub-states. Hence, the
protein becomes biologically active at these temperatures.  This
dynamical transition is common to many biopolymers and is believed to be
triggered by the strong coupling with the mobility of the hydration
water \cite{slaving}, which shows a similar dynamical transition at
about the same $T$ \cite{Chen}. Chen {\it et al.}  by studying the
translational correlation time of water molecules hydrating a lysozyme
protein \cite{Chen}, DNA \cite{ChenDNA06} and RNA \cite{ChenRNA08},
found that at about 220~K the dynamics of hydration water changes from
non-Arrhenius at high $T$ to Arrhenius at low $T$. By definition a
correlation time $\tau$ has an Arrhenius behavior when
\begin{equation}
\tau=\tau_0\exp[E_{\rm A}/k_BT]
\label{eqArr}
\end{equation}
where $\tau_0$ is the correlation time in the high-$T$ limit, $E_{\rm
  A}$ is a $T$-independent activation energy and $k_B$ is the Boltzmann
constant.  On the other hand, $\tau$ is non-Arrhenius when its behavior
cannot be fitted with the expression in Eq.~(\ref{eqArr}).

Motivated by these experiments, Kumar {\it et al.}~\cite{KumarPRL06}
simulated using the TIP5P model the dynamics and thermodynamic behavior
of hydration water  for (i) an orthorhombic form
of hen egg-white lysozyme and (ii) a Dickerson dodecamer DNA at constant
pressure $P=1$~atm, several constant temperatures $T$, and constant
number of water molecules $N$.  Kumar {\it et al.}~\cite{KumarPRL06}
found that the mean square fluctuations $\langle x^2\rangle$ of the
biomolecules changes its functional form below $T_{\rm p}\approx 245$~K,
for both lysozyme and DNA. They also found that the specific heat $C_P$
of the total system (biopolymer and water) displays a maximum at $T_{\rm
  W}\approx (250\pm10)$~K for both biomolecules.

To describe the quantitative changes in structure of hydration water,
Kumar {\it et al.} \cite{KumarPRL06} calculated the local tetrahedral
order parameter $Q$ \cite{Er01} for hydration water surrounding lysozyme
and DNA and found that the rate of increase of $Q$ has a maximum at
$T_{\rm Q}=(245\pm10)$~K, the same temperature of the crossover in the
behavior of mean square fluctuations.  Finally, they found that the
diffusivity of hydration water exhibits a dynamic crossover from
non-Arrhenius to Arrhenius behavior at the crossover temperature
$T_\times\approx (245\pm10)$~K for lysozyme and $T_{\times}\approx
(250\pm10)$~K for DNA. Note that $T_\times$ is much higher than the glass
transition temperature, estimated for TIP5P as $T_g=215$K
\cite{Brovchenko}. Thus this crossover is not likely to be related to
the glass transition in water. Therefore, the fact that $T_{\rm p}
\approx T_{\times} \approx T_{\rm W} \approx T_{\rm Q}$ is evidence of
the correlation between the changes in protein fluctuations and the
hydration water thermodynamics and structure.  Before analyzing in more
details this relation, is worth considering the results of experiments
about water confined in nanostructures.

\section{Confined water}

The non-Arrhenius to Arrhenius dynamic crossover has been found also in
water confined in 20\AA~MCM-41 silica pores.  In 2004, indeed, Faraone
{\it et al.} found the crossover for confined water at $T_L\approx
221$~K by studying the structural relaxation time by neutron
scattering~\cite{Faraone}. This result was reinforced by the neutron
magnetic resonance measurements of Mallamace {\it et al.} showing the
crossover at $T_L\approx 225$~K for the self diffusion of confined
water \cite{Mallamace}.

In 2005, Liu {\it et al.} \cite{Liu} showed that, by increasing the
pressure, the dynamic crossover of water confined in 20\AA~MCM-41 silica
pores disappears at pressure between 1200 and 1600 bar. Xu {\it et al.}
\cite{Xu05}, by using simulations of TIP5P, ST2 water and other models
for liquids with the LLCP, showed that the disappearing of the dynamic
crossover of water can be associated to the presence of the LLCP ($C'$).
The simulations for water, indeed, display the same phenomenology of the
experiments, with a crossover in the diffusion coefficient at $P$ below
the LL critical pressure $P<P_{C'}$, and with no crossover at
$P>P_{C'}$. Xu {\it et al.} \cite{Xu05} presented numerical evidences
that the crossover at $P<P_{C'}$ is associated to the structural change
occurring at the temperature of the maximum of the specific heat
$C_P^{\rm max}$ along a line departing from the LLCP and extending in
the one-phase region of the $P$--$T$ phase diagram. This line, close to
the LLCP, coincides with the Widom line \cite{fs2007}, defined as the
locus of the maximum correlation length in the one-phase region.

Xu {\it et al.} \cite{Xu05} interpreted the absence of the crossover in
the diffusion coefficient at $P>P_{C'}$ as a consequence of the fact
that the HDL-to-LDL spinodal occurs at almost constant $P\simeq P_{C'}$
for decreasing $T$. Therefore, when cooled at $P>P_{C'}$, water never
crosses the HDL-to-LDL spinodal and does not undergo the structural
change responsible for the dynamic crossover.

Hence, although the origin of the crossover has different
interpretations \cite{PRLcomments1}, the experimental and numerical
evidences suggest that the change in the dynamics is triggered
by a local rearrangement of the H bond network
\cite{PRLcomments2}. Experiments and simulations, however, cannot give a
definitive answer due to their finite resolution. For this reason is
interesting to analyze the dynamic crossover of supercooled water within
the framework of a Hamiltonian cell model that allows simulations and
analytic calculations \cite{fsPhysA,fs2007}.

\section{Water model}

We consider a cell model for water \cite{fs,fsPhysA,fms,fs2007} based on
the experimental observations that on decreasing $P$ at constant $T$, or
on decreasing $T$ at constant $P$, (i) water displays an increasing
local tetrahedrality \cite{Darrigo81,Angell84}, (ii) the volume per
molecule increases at sufficiently low $P$ or $T$, and (iii) the O-O-O
angular correlation increases \cite{Soper-Ricci00,ricci}, consistent
with simulations \cite{SGGabinition00,raiteri}.

The system is divided into cells $i\in[1,\ldots,N]$ on a regular
lattice, each containing a molecule, with a volume $v_i\geq v_0$, where
$v_0$ is the hard-core volume of one molecule, with a total volume
$V=\sum_i^N v_i$.  In $d$ dimensions the distance between two nearest
neighbor (n.n.) molecules $i$ and $j$ is $r_{i,j}\equiv
(v_i^{1/d}+v_j^{1/d})/2$. Since $v_i$ is a continuous variable, the
distance $r_{i,j}$ is continuous.

Every cell is occupied by a molecule.  The dimensionless density for the
molecule in cell $i$ is $v_0/v_i\in(0,1]$.  We use a discrete two-state
  liquid-index $n_i$ to quantify if the cell $i$ is in the liquid phase
  or not, with $n_i=1$ if $v_0/v\geq 0.5$ and $n_i=0$
  otherwise. Therefore, $\sum_i^N n_i$ is the total number of liquid
  cells and $N-\sum_i^N n_i$ is the total number of gas cells.

The van der Waals attraction between the molecules is represented by the
Hamiltonian term
\begin{equation} 
{\cal H}\equiv -\epsilon\sum_{\langle i,j\rangle}{n_in_j}~, 
\label{LG} 
\end{equation} 
where $\epsilon>0$ is the van der Waals attraction energy, which induces
the liquid-gas phase transition.

Each molecule $i$ has four H-bond indices $\sigma_{ij} \in
[1,\ldots,q]$, corresponding to four n.n. cells $j$, giving rise to
$q^4$ different molecular orientations.  Bonding and intramolecular (IM)
interactions are accounted for by, respectively, the two Hamiltonian
terms
\begin{equation}
{\cal H}_{\rm B}
\equiv -J\sum_{\langle i,j\rangle}{n_in_j}
\delta_{\sigma_{ij}\sigma_{ji}} ,
\label{HB}
\end{equation}
where the sum is over n.n. cells, $0<J<\epsilon$ is the bond energy,
$\delta_{a,b}=1$ if $a=b$ and $\delta_{a,b}=0$ otherwise, and
\begin{equation}
{\cal H}_{\rm IM}\equiv -J_{\sigma} \sum_{i} n_i \sum_{(k,\ell)_i}
\delta_{\sigma_{ik}\sigma_{i\ell}},
\label{HIM}
\end{equation}
where $\sum_{(k,\ell)_i}$ denotes the sum over the IM bond indices
$(k,l)$ of the molecule $i$ and $J_\sigma>0$ is the IM interaction
energy with $J_{\sigma}<J$, which models the angular correlation between
the bonds on the same molecule.

When H bonds are formed, a small volume $v_{\rm B}$ is added to the
local volumes $v_i$ and $v_j$ of the two H-bonded molecules $i$ and $j$,
increasing their average distance to $r_{ij}\equiv(v_i+v_j+v_{\rm
  B})^{1/d}/2$.  Pictorially, this can be viewed as a local increase of the
excluded volume associated with molecules $i$ and $j$ and is consistent
with the experimental observation that H-bonded molecules form a low
density open structure. Therefore, the total volume is proportional to
the total number $N_{\rm B}$ of H bonds, as
\begin{equation}
V\equiv V_0 + N_{\rm B} v_{\rm B} ~,
\label{vol}
\end{equation}
where $V_0\equiv N v_0$ 
is the volume of the liquid with no H bonds, and 
\begin{equation}
N_{\rm B}\equiv
\sum_{\langle i,j\rangle}{n_in_j\delta_{\sigma_{ij},\sigma_{ji}}} ~.
\label{Nhb}
\end{equation}

Therefore, the cell model has the total Hamiltonian
\begin{equation}
{\cal H}_{\rm tot}\equiv{\cal H}+{\cal H}_{\rm B}+{\cal H}_{\rm IM}
\label{Ham}
\end{equation}
with the volume given by Eq.~(\ref{vol}).

\section{Thermodynamics of the water model}

The cell model described in the previous section has been analyzed by
mean field calculations \cite{fs,fsPhysA,fms,fs2007,kfsPRL,jpcm_kfs} and
Monte Carlo simulations \cite{fms,kfsPRL,jpcm_kfs}.  The model's
parameters considered in these studies are: $J/\epsilon=0.5$,
$J_\sigma/\epsilon=0.05$, $v_{\rm B}/v_0=0.5$ and $q=6$.

\subsection{Mean field}

The mean field results \cite{fs2007}, consistent with computer
simulations \cite{fms}, reproduce the known phase diagram of fluid
water, with the liquid-gas coexistence region ending in the critical
point $C$, at $k_BT_C/\epsilon=1.03\pm 0.03$ and
$P_Cv_0/\epsilon=0.18\pm 0.04$, and with the temperatures of maximum
density (TMD) at constant $P$ that decreases with increasing $P$ as in
the experiments \cite{ANGELL,Poole97}.

In the deeply supercooled region the density has another discontinuity
marking the coexistence region between two liquids at different
densities.  This discontinuity is associated with a discontinuity in
$m_\sigma^{\rm min}$, the value of the tetrahedral order parameter
$m_\sigma\in[0,1]$ that minimizes the molar Gibbs free energy $g$.
Here, $m_\sigma$ quantifies the number of H bonds with tetrahedral
orientation and is defined as the difference between the number-density
of $\sigma_{ij}=1$ and the average number-density for the other $q-1$
states, i.e.,  $m_\sigma\equiv n_\sigma-(1-n_\sigma)/(q-1)$, where
\begin{equation}
n_\sigma \equiv \frac{1}{4nN}\sum_{\langle i,j \rangle}{n_in_j}
\delta_{\sigma_{ij},1}
\end{equation}
and where the state $\sigma_{ij}=1$ corresponds to the appropriate state
to form a (not bifurcated) H-bond with local tetrahedral order.

In the $NPT$ ensemble, the relevant free energy is the Gibbs energy per
mole
\begin{equation}
g\equiv u-Ts+Pv ~, 
\label{gibbs}
\end{equation}
where $u$ is the molar energy, $s$ the molar entropy, and $v$ the molar
volume.  As explained in Ref.~\cite{fs2007}, these quantities can all be
written in the mean field approximation, giving rise to a mean field
expression for $g$ that can be minimized with respect to the gas-liquid
order parameter $m$ and the tetrahedral order parameter $m_\sigma$.

\begin{figure}
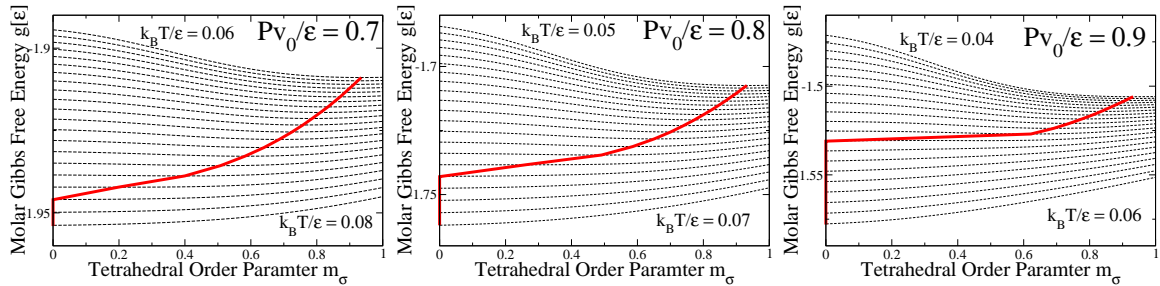

\includegraphics[width=5cm]{franzese_lund_1.eps}
\includegraphics[width=5cm]{franzese_lund_2.eps}
\includegraphics[width=5cm]{franzese_lund_3.eps}
\caption{The mean field molar Gibbs free energy $g$ (dashed lines), in
  units of $\epsilon$, as function of the dimensionless tetrahedral
  order parameter $m_\sigma\in [0,1]$ for different choices of $T$ and
  $P$. The thick (red) line connects the points $m_\sigma^{\rm min}$ of
  minimum $g$ at different $T$ for $Pv_0/\epsilon=0.7$ (left panel), 0.8
  (center panel), 0.9 (right panel). In each panel the topmost line
  corresponds to the lowest $k_BT/\epsilon$ (0.06, 0.05 and 0.04,
  respectively) and the bottom line to the highest $k_BT/\epsilon$
  (0.08, 0.07, 0.06, respectively) with lines separated by $k_B \delta
  T/\epsilon=0.001$. In all the panels $m_\sigma^{\rm min}$ increases
  when $T$ decreases, being 0 at the higher temperatures and $\simeq
  0.9$ at the lowest temperature.  The value $m_\sigma=0$ corresponds to
  the absence of tetrahedral order, i.e. to the high-density arrangement
  of water molecules.  The value $m_\sigma=1$ corresponds to full
  tetrahedral order of the H bonds, i.e. to the low-density arrangement
  of the water molecules.  At $Pv_0/\epsilon=0.7$ (left panel), by
  increasing $T$, $m_\sigma^{\rm min}$ changes without discontinuity
  from $\simeq 0.9$ at $k_BT/\epsilon=0.06$ to 0 at
  $k_BT/\epsilon=0.078$, denoting a continuous change from the
  low-density arrangement at low $T$ to the high-density arrangement at
  high $T$.  At $Pv_0/\epsilon=0.9$ (right panel), instead, by
  increasing $T$, $m_\sigma^{\rm min}$ changes with discontinuity from
  $\simeq 0.6$ at $k_BT/\epsilon=0.051$ to 0 at $k_BT/\epsilon=0.052$,
  denoting a discontinuous phase change from the low-density liquid
  (LDL) to the high-density liquid (HDL). The pressure
  $Pv_0/\epsilon=0.8$ (center panel) is very close to the critical
  pressure $P_{C'}$, below which the discontinuity seen at higher $P$
  disappears.  }
\label{g}
\end{figure}

For any $P$ at low $k_BT/\epsilon<0.1$, $g$ has its minimum for $m=1$
(liquid-phase value) and for a value $m_\sigma^{\rm min}$ that changes
with $T$ and $P$ [Fig.(\ref{g})].  At constant $P$, $m_\sigma^{\rm min}$
decreases with increasing $T$, displaying a discontinuity above
$Pv_0/\epsilon\simeq 0.8$ [Fig.(\ref{g}) right]. The discontinuity
disappears at $Pv_0/\epsilon\simeq 0.8$ [Fig.(\ref{g}) center] and the
approach of $m_\sigma^{\rm min}$ to 0 is always continuous below
$Pv_0/\epsilon= 0.8$ [Fig.(\ref{g}) left].  The appearance of the
discontinuity in $m_\sigma^{\rm min}$ denotes the occurrence at high $P$
of a phase transition between two liquid phases with a different value
of the tetrahedral order parameter $m_\sigma$: an orientationally
disordered phase ($m_\sigma=0$) at high $T$, with no tetrahedral order,
and a tetrahedrally ordered phase ($m_\sigma> 0.5$) at low $T$.  The
phase separation disappears at the liquid-liquid critical point $C'$,
that in mean field is estimated at $k_BT_{C'}/\epsilon= 0.062\pm 0.02$
and $P_{C'}v_0/\epsilon=0.82\pm 0.02$ \cite{fs2007}. The discontinuity
in $m_\sigma$ determines the discontinuity in the density between the
two phases, as can be shown analytically \cite{fs2007}, separating the
liquid in LDL at low $T$ and HDL at high $T$, consistent with the LLCP
scenario.

\subsection{Monte Carlo}

In the Monte Carlo simulations of the cell model, the Hamiltonian ${\cal
  H}_{\rm tot}$ in Eq.(\ref{Ham}) is simulated with the standard
spin-flip method \cite{fms,kfsPRL,jpcm_kfs}. 
For sake of simplicity, we solve the model under the condition of being
in a homogeneous phase with $v_i=v$ for any $i$. Although this condition
can be easily removed, the solution of the model under this condition
gives good qualitative agreement with the experiments, as we will see in
the following sections.

The model is defined in any dimension. However, since we assume for
simplicity that any molecule can form at most four H bonds, we solve the
model on a regular square lattice of cells.  
This choice is particularly appropriate for the study of quasi-2D
water \cite{kbsgs05}
between hydrophobic surfaces and is a first order approximation
to the layer of water between the surface of a protein and the surface
of the frozen bulk water at low $T$. Of course, in this approximation
the interaction with the confined surfaces is not taken into account
\cite{ksbs07}. Nevertheless, the qualitative comparison with the
experiments is satisfactory. 
The lack of the third dimension, and
the number of neighbors limited to four, could affect properties related
to the amount of free space around a molecule, such as the diffusion
constant \cite{girardi,dlSF,hps08}. These properties are not considered in the
analysis of the model reported here.

To further simplify the simulations, the
van der Waals interaction, represented in mean field by the 
Hamiltonian term in Eq.(\ref{LG}), is replaced by an equivalent Lennard-Jones
potential with characteristic energy $-\epsilon$ at distance $R_0\equiv
v_0^{1/d}$ and truncated with a hard-core in its minimum at $R_0$
\cite{fms,kfsPRL,jpcm_kfs}. Simulations performed with a standard
Lennard-Jones potential, without truncation, show that the results are
qualitatively unchanged \cite{dlSF}.

\begin{figure}
\includegraphics[width=9cm]{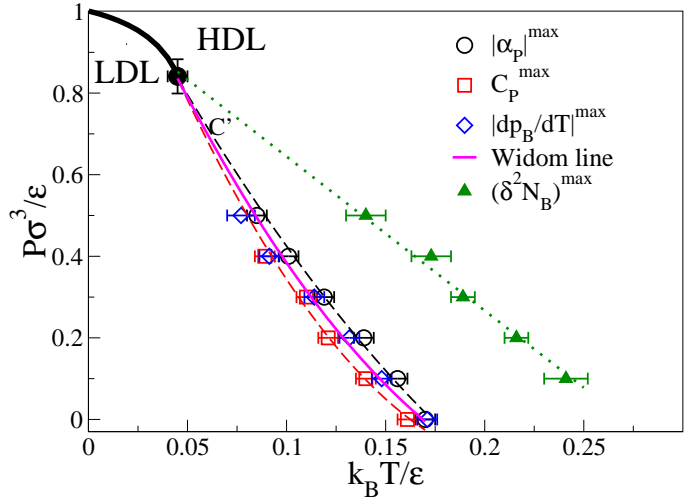}
\caption{ The Monte Carlo phase diagram for the cell model at low $T$
  for $N=3600$ water molecules \cite{jpcm_kfs}.  $C'$ is the HDL-LDL
  critical point, end of first-order phase transition line (thick line)
  \cite{fs}; symbols are maxima of the coefficient of thermal expansion
  $|\alpha_P|^{\rm max}$ ($\bigcirc$), isobaric specific heat $C_P^{\rm
    max}$ ($\Box$), $|dp_{\rm B}/dT|^{\rm max}$ ($\Diamond$) the
  numerical derivative of the probability of forming a H-bond,
  proportional to the fluctuation of the number of bonds $(\delta^2
  N_{\rm B})^{\rm max}$ ($\triangle$); the Widom (solid) line,
  corresponding to locus of maximum correlation length and estimated as
  the average between $|\alpha_P|^{\rm max}$ and $C_P^{\rm max}$,
  coincides within the error bars with $|dp_{\rm B}/dT|^{\rm max}$,
  i.e. with the locus of the maximum structural variation.  Dashed lines
  are guide for the eyes.}
\label{figMC}
\end{figure}

The resulting phase diagram \cite{fms} displays, at high $T$ and low
$P$, a gas-liquid phase transition ending in a gas-liquid critical point
$C$. From $C$ departs in the supercritical region the gas-liquid Widom
line, i.e. the line of maximum ---but finite--- correlation length for
the fluid.  At lower $T$, the phase diagram displays the TMD line,
retracing toward low $T$ at high $P$ as in the experiments
\cite{debenedettibook}.  At lower $T$ and high $P$, the phase diagram
displays a HDL-LDL phase transition with negative slope in the $P$--$T$
plane, ending in a LL critical point $C'$.  As for the gas-liquid
critical point, also from $C'$ the line of maximum correlation length
for the liquid, the LL Widom line \cite{Xu05,fs2007}, departs into the
one-phase region (Fig.~\ref{figMC}).  Therefore, the Monte Carlo results
confirm the mean field analysis, displaying a phase diagram consistent
with the LLCP scenario.

Recent calculations \cite{kevin-condmat} show that by varying the
parameters of the cell model, it is possible to recover also ($i$) the
singularity free scenario, ($ii$) the stability limit scenario and
($iii$) the critical point free scenario. In the following we report
about the analysis of the dynamics of the model for the LLCP scenario
and the singularity free (SF) scenario.

\section{Dynamics of the water model}

\begin{figure}
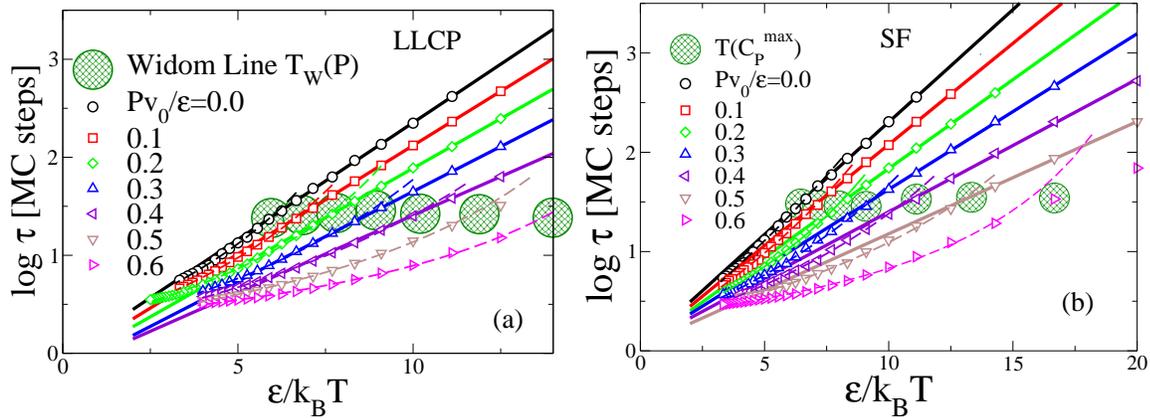

\includegraphics[width=7.5cm]{franzese_lund_5.eps}
\includegraphics[width=7.5cm]{franzese_lund_6.eps}
\caption{Monte Carlo results for the dynamic crossover in the
  orientational relaxation time $\tau$ of the cell water model for a
  range of different pressures \cite{kfsPRL}. (a) In the case of the
  liquid-liquid critical point (LLCP) scenario , the crossover occurs at
  a temperature consistent with the Widom line $T_W(P)$.  (b) In the
  case of the singularity free (SF) scenario, the crossover occurs at a
  temperature consistent with the $T(C_P^{\rm max})$. In both panels the
  large hatched circles mark $T_W(P)$ and $T(C_P^{\rm max})$, with a
  radius approximately equal to the error bar.  Solid and dashed lines
  represent Arrhenius and Vogel-Fulcher-Tamman fits, respectively.  In
  both cases, Kumar {\it et al.} \cite{kfsPRL} predicted that the
  dynamic crossover occurs at approximately the same value of $\tau$ for
  all the values of pressures studied, i.e. the crossover is {\it
    isochronic}.  }
\label{tau}
\end{figure}

We first consider the case of the LLCP scenario.  Performing Monte Carlo
simulations for $T$ and $P$ around the values of the LL Widom line
$T_W(P)$, Kumar {\it et al.} \cite{kfsPRL} found that the correlation
time $\tau$ of $S_i\equiv\sum_j\sigma_{ij}/4$, which quantifies the
degree of total bond ordering for site $i$, displays a dynamic crossover
from Vogel-Fulcher-Tamman (VFT) function at high $T$ to Arrhenius
$T$-dependence at low $T$ (Fig.~\ref{tau}.a).  The VFT function is given
by
\begin{equation}
\tau^{\rm VFT}\equiv\tau_0^{\rm VFT} \exp\left[{T_1\over T-T_0}\right],
\label{VFT}
\end{equation}
where $\tau_0^{\rm VFT}$, $T_1$, and $T_0$ are three fitting parameters,
and the Arrhenius function is given in Eq.(\ref{eqArr}).  This result is
qualitatively consistent with the dynamic crossover found in experiments
for confined water and hydration water
\cite{Chen,Mallamace,ChenDNA06,ChenRNA08,Faraone,Liu} and has been
related to the presence of the LLCP \cite{Xu05,KumarPRL06}.

It is, therefore, interesting to compare this result with the dynamic
behavior of the cell model in the hypothesis in which the SF scenario
holds. To do this, Kumar {\it et al.} \cite{kfsPRL} analyzed the
dynamics of the cell model when the H-bond correlation is zero
($J_\sigma=0$) and the cell model recovers the SF model of Sastry {\it
  et al.} \cite{Sastry96}.

The result (Fig.~\ref{tau}.b) \cite{kfsPRL} shows that also in this
scenario a dynamic crossover is expected. The temperature of the
crossover is coinciding, within the numerical precision, with the
temperature $T(C_P^{\rm max})$ of maximum isobaric specific heat, which
in the SF scenario play a role equivalent to $T_W(P)$ of the LLCP
scenario. Indeed, both $T(C_P^{\rm max})$ and $T_W(P)$ mark the
temperature $T_{\rm struct}^{\rm max}$ of the maximum structural change
for the liquid \cite{jpcm_kfs}.

The cell model allows to clarify that the dynamic crossover is, indeed,
a direct consequence of the structural change occurring at $T_{\rm
  struct}^{\rm max}$. By calculating, in the mean field approximation,
the $T$-dependent activation energy $E_{\rm A}(T)^{\rm MF}$ necessary to
($i$) break a non-tetrahedral H bond, ($ii$) reorient the water molecule
and ($iii$) form a new H bond in a tetrahedral orientation, Kumar {\it
  et al.} \cite{jpcm_kfs} calculated the mean field correlation time as
\begin{equation}
\tau^{\rm  MF}=\tau_0\exp\left[{E_{\rm A}(T)^{\rm  MF} \over k_BT}\right],
\label{ArrMF}
\end{equation}
and compared $\tau^{\rm MF}$ with the correlation time from the Monte
Carlo simulations, finding an excellent agreement.  Since $E_{\rm
  A}^{\rm MF}$ is an explicit function of the number $N_{\rm B}$ of H
bonds, therefore $\tau^{\rm MF}$ is a function of the structure of the
liquid \cite{jpcm_kfs}. Since $N_{\rm B}$ rapidly changes for $T>T_{\rm
  struct}^{\rm max}$, has its maximum variation at $T_{\rm struct}^{\rm
  max}$, and slowly changes for $T<T_{\rm struct}^{\rm max}$, the
correlation time is a non-Arrhenius function of $T$ for $T>T_{\rm
  struct}^{\rm max}$ and almost an Arrhenius function for $T<T_{\rm
  struct}^{\rm max}$, giving rise to a dynamic crossover
\cite{jpcm_kfs}. The agreement between the mean field calculation and
the Monte Carlo results clarifies that the low-$T$ regime is only
approximately Arrhenius.

The analysis of the cell model allows, furthermore, to make a number of
predictions about the dynamics of water at low $T$.

\begin{itemize}

\item[(i)] At the dynamic crossover the value of $\tau$ is about the
  same for all the pressures considered in Ref.~\cite{kfsPRL}, i.e. the
  crossover is {\it isochronic} and the correlation time at the
  crossover is $\tau_{\rm L}=10^{3/2}$ in units of Monte Carlo steps
  \cite{kfsPRL}. Comparison with experiments, in the next section,
  allows to convert this result in real time units.

\item[(ii)] The activation energy $E_{\rm A}$, that is found by fitting
  the low-$T$ dynamic regime with an Arrhenius behavior, decreases
  linearly for increasing $P$ \cite{kfsPRL}.

\item[(iii)] If we fix a characteristic time, for example the crossover
  time $\tau_{\rm L}$, the temperature $T_{\rm L}$ at which this
  correlation time is reached decreases linearly for increasing $P$
  \cite{kfsPRL}.

\end{itemize}

All these predictions are verified in both the LLCP scenario and the SF
scenario. However, Kumar {\it et al.} \cite{kfsPRL} found a difference
between the two scenarios: for the LLCP scenario the index $E_{\rm
  A}/(k_BT_{\rm L})$ increases for increasing $P$, while for the SF
scenario $E_{\rm A}/(k_BT_{\rm L})$ is constant. However, the predicted
increase of $E_{\rm A}/(k_BT_{\rm L})$ is of the order of 1\%
\cite{kfsPRL}. As we will discuss in the next section, this increase is
within the present experimental error bar.

\section{Comparison with the experiments on hydration water}

Recently, Chen and coworkers have performed extensive QENS experiments
to investigate the dynamical behavior of hydration water on lysozyme,
spanning a range of pressures going from ambient pressure up to 1600~bar
\cite{Chu2008}. By measuring the mean square displacement of the of
H-atoms in lysozyme, they found a dynamic crossover for all pressures
studied \cite{Chu2008}.  They also found that the translational
correlation time of the H-atoms of the hydration water shows a dynamic
crossover for $P=1$, 400, 800, 1200, 1500~bar \cite{Chu2008}.  For
$P\leq 1500$~bar, they showed that the lysozyme dynamic crossover and
the hydration water dynamic crossover occurs at the same $T$
\cite{Chu2008}, extending their previous results on lysozyme
\cite{Chen}, DNA \cite{ChenDNA06} and RNA \cite{ChenRNA08} at
ambient pressure.

They also showed that  the dynamic crossover for the
hydration water disappears at $P=1600$~bar \cite{Chu2008}. This
finding, together with 
their previous measurements for water confined in MCM-41 Silica pore
\cite{Liu}, suggests that the disappearing of the non-Arrhenius to
Arrhenius crossover at $P=1600$~bar is independent of the constraint
used to avoid homogeneous ice nucleation.  As shown in Ref.~\cite{Xu05},
the disappearing of the crossover could be the consequence of the LLCP
occurring at a pressure between 1500 and 1600~bar.

The comparison of the experimental data in Ref.~\cite{Chu2008} with the
results of the Monte Carlo simulations allows several observations.

\begin{figure}
\includegraphics[width=9cm]{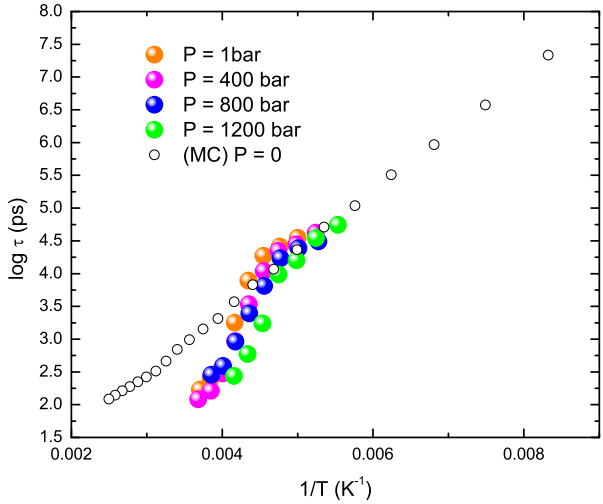}
\caption{Comparison between the Monte Carlo \cite{kfsPRL} results and
  the QENS data \cite{Chu2008} for the correlation time $\tau$ of
  supercooled water. Monte Carlo results (open symbols) are for $P=0$
  and are rescaled in such a way that the crossover occurs at at the
  same $T$ and $\tau$ of the experimental data (full symbols). The
  experimental data are for pressures going from ambient to 1200~bar.  }
\label{compTau}
\end{figure}

\begin{figure}
\includegraphics[width=7.2cm]{franzese_lund_8.eps}
\hfill
\includegraphics[width=7cm]{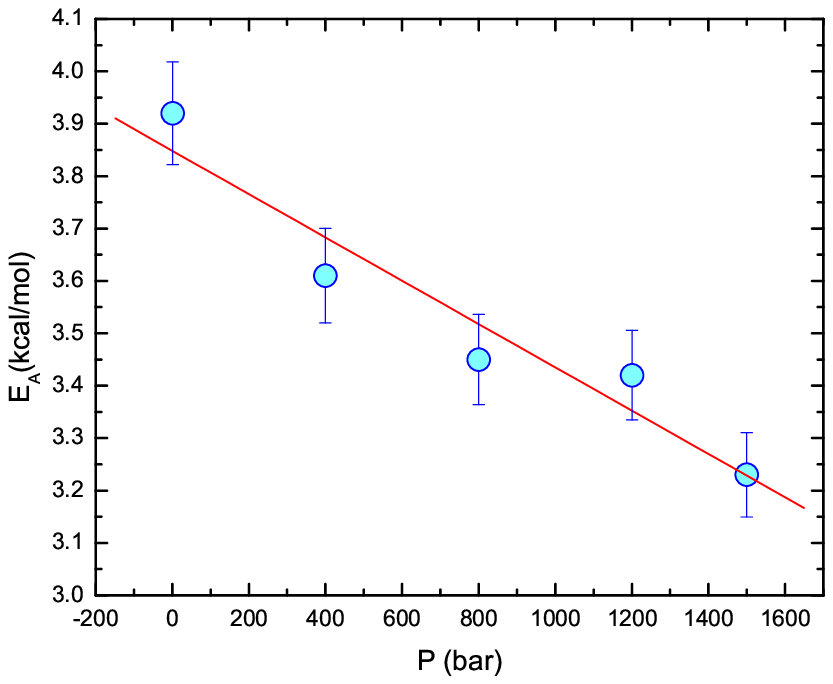}
\includegraphics[width=7.5cm]{franzese_lund_10.eps}
\hfill
\includegraphics[width=7cm]{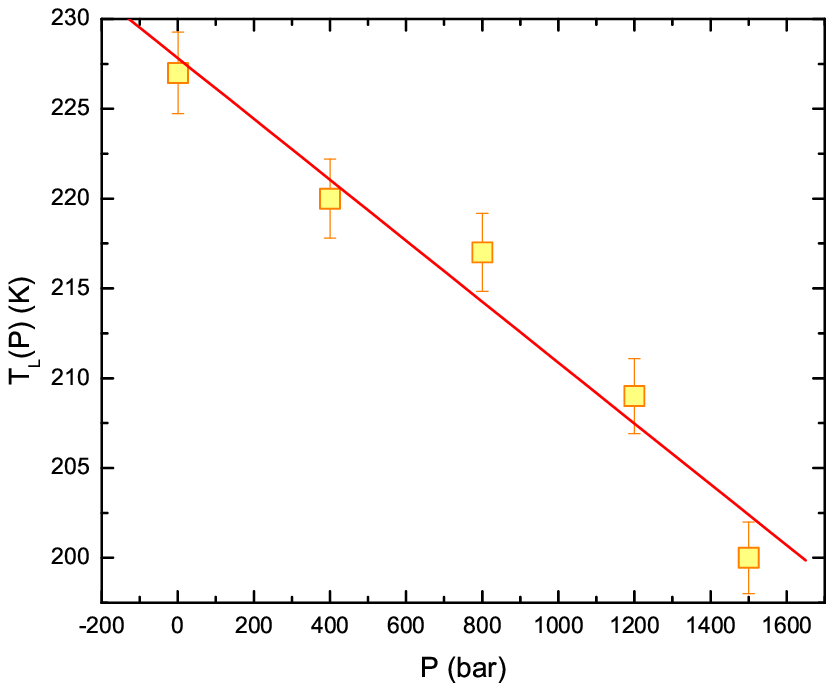}
\caption{Monte Carlo \cite{kfsPRL} results and the QENS data
  \cite{Chu2008} for the activation energy $E_{\rm A}$ of the low-$T$
  regime and the crossover temperature $T_{\rm L}$, as functions of the
  pressure $P$.  Upper left panel: Monte Carlo results for $E_{\rm A}$
  \cite{kfsPRL}; for sake of comparison with the experimental data,
  pressure is rescaled by an arbitrary factor
  $1500$~bar$/(0.8\epsilon/v_0)$ and energy by a factor
  16.4~kcal~mol$^{-1}/\epsilon$ $\simeq$ 68.7~kJ~mol$^{-1}/\epsilon$;
  both the liquid-liquid critical point (LLCP) scenario results
  (circles) and the singularity free (SF) scenario results (squares) are
  reported.  Lower left panel: Monte Carlo results for the crossover
  temperature $T_{\rm L}$; pressure is rescaled as above and temperature
  is rescaled by a factor 220~K$/(0.16\epsilon/k_B)$; symbols are as in
  previous panel.  Upper right panel and lower right panel: QENS data
  for the lysozyme hydration water \cite{Chu2008}.  }
\label{compEaTl}
\end{figure}

\begin{itemize}

\item The correlation time $\tau_{\rm L}$ at the crossover is constant
  as predicted by the cell model, point (i) in the previous section.

\item $\tau_{\rm L}=10^{3/2}$ in units of Monte Carlo steps (MCS)
  \cite{kfsPRL} corresponds to $10^{4.25}$~ps, i.e. 1~MCS$\approx
  600$~ps (Fig.~\ref{compTau}).

\item The high-$T$ (non-Arrhenius) behavior of the correlation time in
  real water has a stronger dependence on $T$ than the one seen in the
  cell model at $P=0$ (Fig.~\ref{compTau}) and is more similar to the
  behavior seen in the model at higher $P$. Since in the model this
  behavior is regulated by the increase of number $N_{\rm B}$ of H bonds
  for decreasing $T$, this observation implies that in real water
  $N_{\rm B}$ rapidly increases approaching the temperature of maximum
  structural change $T_{\rm struct}^{\rm max}$, and $N_{\rm B}$ is
  smaller in real water than in the model for $T>T_{\rm struct}^{\rm
    max}$ \cite{jpcm_kfs}.

\item The low-$T$ (approximately Arrhenius) behavior is well reproduced
  by the cell model (Fig.~\ref{compTau}). This result implies that the
  estimate of the activation energy at $P=0$ in the model compares with
  a reasonable agreement to the activation energy measured for hydration
  water.

\item The direct comparison of the Monte Carlo estimated activation
  energy $E_{\rm A}$ and the QENS data (Fig.~\ref{compEaTl}, upper
  panels) shows that the prediction (ii) of the previous section is
  verified in the experiments, with $E_{\rm A}$ decreasing linearly for
  increasing $P$. The agreement is only qualitative, since the variation
  of $E_{\rm A}$ in the cell model is too large with respect to the
  experiments.

\item The crossover temperature $k_BT_{\rm L}$ in the experiments
  verifies the prediction (iii) of the cell model, i.e. $k_BT_{\rm L}$
  decreases linearly for increasing $P$ (Fig.~\ref{compEaTl}, lower
  panels). Also in this case, the variation in the model is too large
  and the agreement is only qualitative.

\item The crossover temperature $k_BT_{\rm L}/\epsilon\simeq 0.16$ at
  ambient pressure corresponds to 220~K (Fig.~\ref{compTau}), leading to
  a van der Waals interaction strength $\epsilon\simeq 11$~kJ~mol$^{-1}$
  and a H-bond strength $J=0.5 \epsilon=5.5$~kJ~mol$^{-1}$. This H-bond
  strength is about 1/4 of the H-bond strength expected below
  100$^\circ$~C \cite{Khan2000} and is about the value estimated for
  then van der Waals attraction based on isoelectronic molecules at
  optimal separation \cite{Henry2002}.

\end{itemize}

We finally observe that the prediction of the model about the different
behavior of the index $E_{\rm A}/(k_BT_{\rm L})$ in the two scenarios,
cannot be verified in the experiments, since the predicted difference
between the two scenarios (of the order of 1\%) is within the error bars
of the measurements \cite{Chu2008}. The QENS data in Ref.~\cite{Chu2008}
show a non-monotonic behavior for this index within an error bar larger
than the 1\% variation predicted for the LLCP scenario. Hence, the
data could be considered also constant within the error bars, as
predicted for the SF scenario. Therefore, is not possible to 
discriminate between the two scenarios on the basis of the present data
for the index $E_{\rm A}/(k_BT_{\rm L})$.

\section{Discussion and Conclusions}

Several questions remain open.  The work reported here allows us to
propose possible answers to some of these questions.

\begin{itemize}
 
\item {\it Why does water have anomalies?} 

Simulations \cite{Poole} first and, after, experiments
\cite{Soper-Ricci00} have shown the existence of two local 
configurations for water molecules: tetrahedral up to the second shell
(or open) and non-tetrahedral (or closed). These local arrangements
are the consequence of the orientational character of the H bonds and
their typical competition between attraction and repulsion. Many of
the anomalies can be understood on the basis of the tendency of the
H bond to form open configurations to minimize the energy,
frustrated by the necessity of reducing the occupied volume at high
density or pressure.

\item {\it What are the implications of these open and closed
  configurations?} 

The cell model discussed here \cite{fs,fsPhysA,fms,fs2007} shows that
the existence of these local configurations, related to the tendency of
the H bonds to correlate and order in a tetrahedral way, is enough to
imply the occurrence of a liquid-liquid phase transition, possibly
ending in a critical point. Only in the hypothesis that the H bonds
formed by the same molecule are completely uncorrelated, as in the
Sastry {\it et al.} model \cite{Sastry96}, the open and closed
configurations do not separate into two phases, giving rise to the
singularity free scenario.

\item {\it Why is there a dynamic crossover in the correlation time of
  supercooled water?}

Experiments \cite{Chen} and simulations \cite{Xu05} offer evidences that
the dynamic crossover is due to the local variation of H bond network
\cite{KumarPRL06}. The cell model discussed here, furthermore, offers an
analytic relation, based on a mean field approximation, between the
dynamic crossover and the structural change \cite{kfsPRL,fs2007}.  In
the cell model it is assumed that the relevant mechanism for the low-$T$
water dynamics is the breaking of the H bonds to reorient the molecule
and form more tetrahedral bonds, as seen in simulations
\cite{Laage-Hynes2006}.
  
\item {\it Why is the dynamic crossover at the locus $T(C_P^{\rm max})$
  of maximum specific heat?}

At $T(C_P^{\rm max})$, the number $N_{\rm B}$ of H bonds has its maximum
variation.  The liquid changes from the HDL-like local structure at high
$T$ to LDL-like local structure at low $T$. In the HDL-like regions
$N_{\rm B}$ rapidly increases when $T$ decreases, implying that the
activation energy $E_{\rm A}$ for the breaking and reorientation of the
H bonds increases and the behavior is non-Arrhenius \cite{jpcm_kfs}.  In
the LDL-like regions, $N_{\rm B}$ is almost constant with $T$, implying
that $E_{\rm A}$ is approximately constant and the behavior is
asymptotically Arrhenius \cite{jpcm_kfs}.

\item {\it Is the dynamic crossover an evidence of the liquid-liquid
  critical point?}

The dynamic crossover is consistent with both the liquid-liquid critical
point scenario and the singularity free scenario \cite{kfsPRL}. However,
in the hypothesis of the presence of a liquid-liquid critical point is
possible to rationalize the disappearing of the dynamic crossover at
high pressure \cite{Liu,Xu05,Chu2008}, while in the other scenarios no
change in the crossover is expected with the increase of $P$.

\item {\it Is the dynamic crossover a cooperative process?}

The dynamic crossover is related to the increase of the correlation
length $\xi$ in the liquid. This quantity increases for increasing $P$
along the Widom line and diverges at the liquid-liquid critical
point. Therefore, the dynamic crossover is a cooperative process at the
liquid-liquid critical point.
 
\item {\it How pressure studies can help us in understanding the physics
  of water?}

The recent experiments on protein hydration water \cite{Chu2008} show
that is possible to analyze the dynamic crossover of water at very low
$T$, measuring quantities, such as the crossover temperature $T_{\rm L}$
and the activation energy $E_{\rm A}$, whose behaviors have been
predicted by the theory \cite{kfsPRL}. The experiments have verified
three of the four theoretical predictions, being the forth  within the
error bars.

\end{itemize}

In conclusion, in this report we compare the predictions of a cell
model for water analyzed by MC in 2D \cite{kfsPRL} with the QENS data
for water at the 
surface of lysozyme \cite{Chu2008}. The data show a dynamic crossover
from high-$T$ 
non-Arrhenius behavior to low-$T$ quasi-Arrhenius behavior for the water
relaxation time at $P<1600$~bar. Both the temperature $T_{\rm L}(P)$ and
the low-$T$ average activation energy $E_{\rm A}(P)$ 
linearly decrease for increasing $P$, as predicted by the model. 
The relaxation time at the crossover $\tau(T_{\rm L})$ is independent of
$P$, as in the model. 
It is an open question if this isochronic behavior is related to the
constant-time dynamic crossover observed in glass formers \cite{casalini}.

The mean field approach \cite{kfsPRL,jpcm_kfs} allows to find a
functional relation between $E_{\rm A}$ and 
the number $N_{\rm B}(T,P)$ of H bonds, i.e. allows to show in a
clear way that
the dynamic crossover is a consequence of the structural changes
occurring at the temperature of maximum specific heat 
$T(C_P^{\rm max})$. However, the comparison with the experiments
does not clarify which scenario describes better the supercooled
region of water. Indeed, within the experimental error bar larger
than 1\%, both the SF and the LLCP scenario could be
consistent. Nevertheless, the disappearing of the crossover observed in
the experiments at $P=1600$~bar is expected only in the presence of a LLCP.

\section*{Acknowledgments}

We thank our collaborators, M. I. Marqu\'es, M. Yamada, and F. de los
Santos, and the NSF Chemistry Program CHE~0616489 for support.
G. F. also thanks the Spanish Ministerio de Ciencia e Innovaci\'on,
project FIS2007-61433.

\end{document}